\documentclass[bibyear]{aa}
\usepackage{natbib} 

\usepackage{graphicx}
\usepackage[usenames,dvipsnames,svgnames,table]{xcolor}
\usepackage{booktabs}
\usepackage{multirow}
\usepackage{txfonts}
\usepackage{color}

\begin{document}

   \title{Spatial gradients of GCR protons in the inner heliosphere derived from Ulysses COSPIN/KET and PAMELA measurements}
   \titlerunning{Spatial gradients of GCR protons in the inner heliosphere}

   \subtitle{}

   \author{J. Gieseler
          %\inst{1}
          \and
          B. Heber%\inst{1}
          }

   \institute{Institute of Experimental and Applied Physics, University of Kiel, 
                Leibnizstr. 11, 24118 Kiel, Germany\\
              \email{gieseler@physik.uni-kiel.de}
             }

   \date{}

% \abstract{}{}{}{}{} 
% 5 {} token are mandatory
 
  \abstract
  % context heading (optional)
  % {} leave it empty if necessary  
   {During the transition from solar cycle 23 to 24 from 2006 to 2009, the Sun was in an unusual solar minimum with very low activity over a long period. These exceptional conditions included a very low interplanetary magnetic field (IMF) strength and a high tilt angle, which both play an important role in the modulation of galactic cosmic rays (GCR) in the heliosphere. Thus, the radial and latitudinal gradients of GCRs are very much expected to depend not only on the solar magnetic epoch, but also on the overall modulation level. }
  % aims heading (mandatory)
   {We determine the non-local radial and the latitudinal gradients of protons in the rigidity range from $\sim$0.45  to 2\,GV.}
  % methods heading (mandatory)
   {This was accomplished by using data from the satellite-borne experiment Payload for Antimatter Matter Exploration and Light-nuclei Astrophysics (PAMELA) at Earth and the Kiel Electron Telescope (KET) onboard Ulysses on its highly inclined Keplerian orbit around the Sun with the aphelion at Jupiter's orbit.}
  % results heading (mandatory)
   {In comparison to the previous A$>$0 solar magnetic epoch, we find that the absolute value of the latitudinal gradient 
%is smaller and larger at higher and lower rigiditie
is lower at higher and higher at lower rigidities. %, respectively. 
This energy dependence is therefore a crucial test for models that describe the cosmic ray transport in the inner heliosphere.}
  % conclusions heading (optional), leave it empty if necessary 
   {}


   \maketitle
%
%________________________________________________________________

\section{Introduction}

\begin{figure}
 \centering
% \noindent
% \fbox{\includegraphics[viewport=-47 -27 465 730, clip, width=20pc]{jgr_overview_update_2013.pdf}}
% \fbox{\includegraphics[viewport=0 22 532 773, clip, width=20pc]{jgr_overview_update_2013.pdf}}
 \includegraphics[viewport=2 442 516 771, clip, width=\hsize]{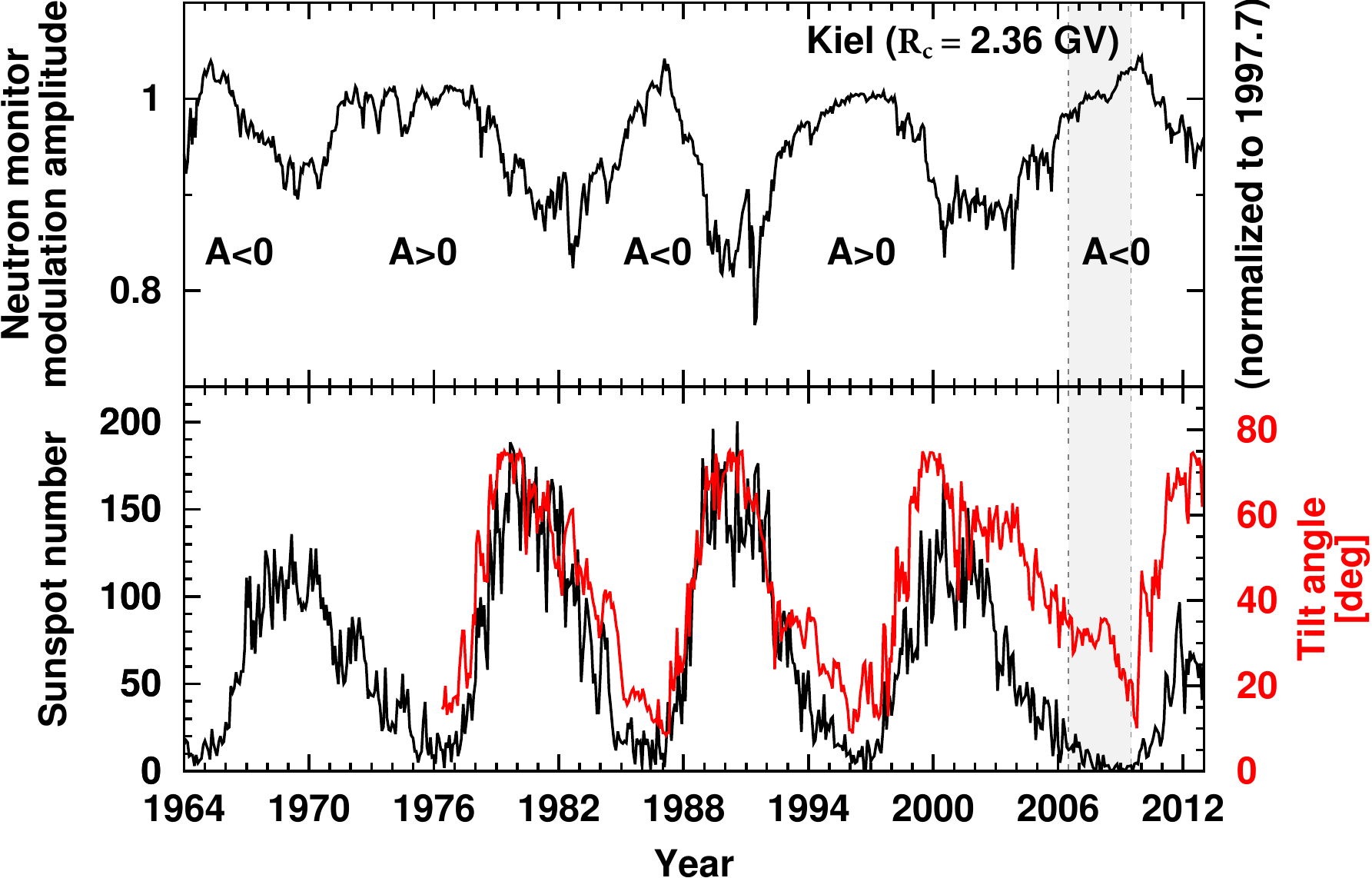}
 \caption{
%jgr overview update 2013 4 paper.eps
Top: Monthly Kiel neutron monitor count rate (cutoff rigidity $R_{c}$~=~2.36\,GV),  
%and calculated force field approximation intensity for 20.92\,GV protons, both 
normalized to 1997.7. Bottom: Monthly sunspot number from the Royal Observatory of Belgium and tilt angle (classic model) from the Wilcox Solar Observatory. The  period investigated in  this work is shaded.
%\textcolor{red}{{Replace the neutron monitor with Ulysses KET P190 and SOHO 250 to above 1 GeV from Patrick. }}
}
 \label{overview}
 \end{figure}

Galactic cosmic rays (GCRs) that propagate in the heliosphere are affected by the solar activity. They are scattered at magnetic field irregularities, undergo convection and adiabatic deceleration in the expanding solar wind, and are exposed to gradient and curvature drifts in the large-scale heliospheric magnetic field. This results in the modulation of GCRs with solar activity, shown in %the top panel of 
Fig.~\ref{overview}. The upper panel displays the (normalized) count rate of the Kiel neutron monitor %as an proxy for GCRs is 
plotted over time. 
A simple comparison with the sunspot number in the panel below shows the anticorrelation between solar activity and GCR intensity. 
In the 1960s, 1980s and 2000, when the solar magnetic field pointed toward the Sun in the northern hemisphere (so-called A$<$0-magnetic epoch), the time profiles were peaked, whereas they were more or less flat in the 1970s and 1990s during the A$>$0-solar magnetic epoch, showing a correlation with the 22-year solar magnetic cycle. 

To understand solar and heliospheric modulation, it is vital to reproduce the spatial distribution of cosmic rays in the three-dimensional heliosphere, that is, around solar minimum periods. Keys to fulfilling this task are measuring the cosmic ray distribution in the three-dimensional heliosphere and modeling the cosmic ray transport. 
An important prediction from drift-dominated modulation models is the expectation that protons  will have large positive and negative latitudinal gradients in an A$>$0- and A$<$0-solar magnetic epoch, respectively. In agreement with expectations, the latitudinal gradients observed in the 1990s were positive \citep{Heber1996a, Heber1996b, Simpson1996}. But in contrast to these expectations, \citet{Heber1996a, Heber-etal-2006} %Heber et al. (\citeyear{Heber1996a, Heber-etal-2006}) 
showed that the measured spectrum over the poles in 1994 and 1995 was still lower than the Voyager measurements at 62~AU and
that it was highly modulated. Not only were the latitudinal gradients much smaller than anticipated, the energy dependence also showed an unexpected maximum at a rigidity of about 1\,GV \citep[see Fig.~8 in][]{Heber1996a}. 

These investigations relied on particle measurements made by the Cosmic Ray and Solar Particle Investigation Kiel Electron Telescope (COSPIN/KET) and High Energy Telescope (COSPIN/HET) using particle measurements from the IMP~8 satellite as a baseline close to Earth. Unfortunately, in 2006 IMP~8 was lost and a new baseline only became available when the Payload for Antimatter Matter Exploration and Light-nuclei Astrophysics (PAMELA) experiment was launched in July 2006. \citet{DeSimone2011} analyzed proton data at $1.6-1.8$\,GV from Ulysses COSPIN/KET and PAMELA for the period from launch of PAMELA in 2006 to the end of Ulysses in 2009. They showed in agreement with the model calculation that the latitudinal gradients were negative, but again in contrast to the prediction, these gradients were much smaller than expected. 

The Voyager~1 spacecraft located beyond 120~AU in the outer heliosheath \citep{Stone-etal-2013, Krimigis-etal-2013} has shown that the local interstellar spectrum for ions is known with a low uncertainty \citep{Potgieter-etal-2014}. These small uncertainties result from the fact that there might even be modulation in the outer heliosheath \citep{Scherer-etal-2011,Herbst-etal-2012,Strauss-etal-2013}. Among others, the rigidity dependence of the latitudinal gradient in the A$<$0-solar magnetic epoch is a crucial quantity that helps to determine the propagation parameters in the heliosphere. 

\section{Instrumentation}
The determination of non-local gradients relies on measurements that are well calibrated against each other. During Ulysses' first fast latitude scan in the 1990s, a baseline close to Earth for the Ulysses COSPIN/KET (see Sect. \ref{KET}) was the University of Chicago instrument onboard the IMP~8 spacecraft. After 2006, the analysis had to rely on measurements of the PAMELA instrument (see Sect. \ref{Pamela}).
\subsection{Ulysses Kiel Electron Telescope \label{KET}} 
Ulysses was a joint ESA/NASA mission that was launched in October 1990 and was switched off in June 2009. During more then 18 years of measurements, the spacecraft performed three of its highly inclined ($80.2^{\circ}$) orbits around the Sun, with the aphelion at Jupiter's orbit and the perihelion close to 1~AU. Part of these orbits were three so-called fast latitude scans, during which the spacecraft covered a latitude range from $-80^{\circ}$ to $+80^{\circ}$ in roughly one year. The three southern polar passes occurred from 1994-06-26 to 1994-11-05, 2000-09-06 to 2001-01-16 and 2006-11-17 to 2007-04-03, respectively. The corresponding passes of the northern polar region were from 1995-06-19 to 1995-09-29, 2001-08-31 to 2001-12-10 and from 2007-11-30 to 2008-03-15. While the second polar pass took place during solar maximum periods, the first and third polar passes were performed during solar minimum conditions during an A$>$0 and A$<$0-solar magnetic epoch, respectively. The time period that we analyze here is shaded in Fig. 1.
%(see bottom panel of Fig.~\ref{overview}) 

The Kiel Electron Telescope (KET) onboard Ulysses was part of the Cosmic Ray and Solar Particle Investigation (COSPIN) experiment and measured electrons, protons, and $\alpha$-particles in the range from 2.5~MeV to above 300~MeV and from 4~MeV/n to above 2~GeV/n, respectively \citep[see][]{Simpson1992}.
%
%%%%%%%%%%%%%%%%%%%%%%%%%%%%%%%%%%%%%%%%%%%
%%%%%%%%%%%%%%%%%%%%%%%%%%%%%%%%%%%%%%%%%%%
%
\subsection{PAMELA \label{Pamela}}
The ongoing experiment  called Payload for Antimatter Matter Exploration
and Light-nuclei Astrophysics (PAMELA) is a spectrometer onboard a Russian Resurs-DK1  satellite, launched on a polar elliptical orbit around Earth in June 2006 \citep{Picozza2007}. 
Its main purpose is the measurement of electrons, positrons, protons, antiprotons, and light nuclei over a very wide range of energy up to hundreds of GeV.
With an altitude between 350~km and 600~km, the detection of charged particles down to 50~MeV is only possible during high-latitude phases with a low geomagnetic cutoff.

\section{Gradient calculation}
\begin{figure}
 \centering
%  \noindent
  \includegraphics[viewport=7 15 600 399,clip, width=\hsize, angle=0]{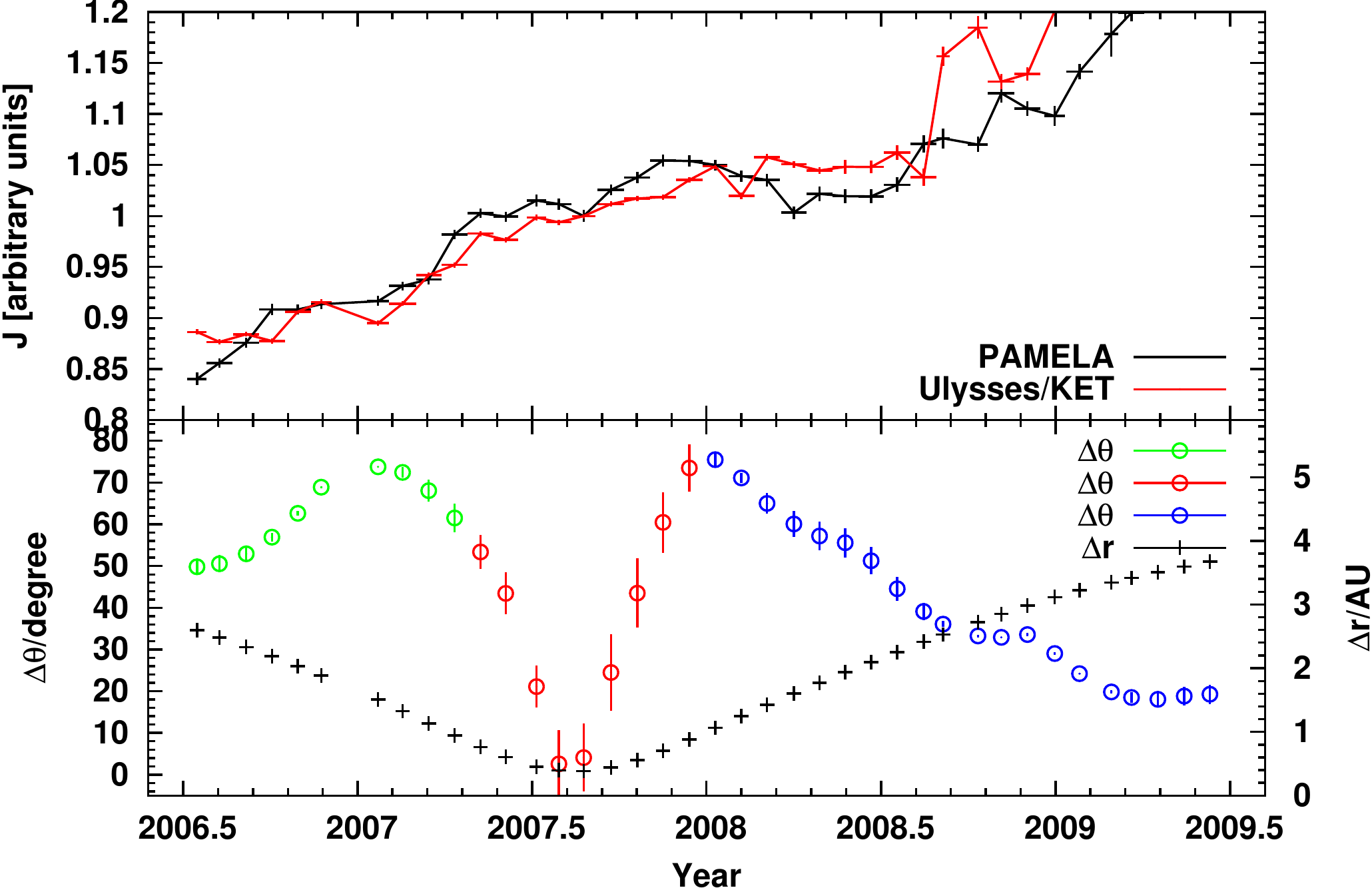}
  \caption{Top: Intensities of $\sim$1.9\,GV protons measured by Ulysses/KET and PAMELA, normalized in August 2007 (see Sect.~\ref{subsection_normalization}). Bottom: Differences in radial and latitudinal distance ($\Delta r$ and $\Delta\theta$) between Ulysses and PAMELA for the analysis period. The fast latitude scan of Ulysses is plotted in red, green and blue indicate its slow ascent and descent, respectively.
%Indicated by shadings are the two time periods $t_1$ and $t_2$ used for the analysis of temporal evolution at Earth and Ulysses. The intensity ratios of these periods are shown in Fig.~\ref{all_ratio_paper_rig}
%where Ulysses was at nearly the same radial distance ($r_1 \approx r_2$) and absolute latitude ($|\theta_1| \approx |\theta_2|$). 
}
  \label{grad_orbit}
  \end{figure}

To calculate the rigidity-dependent \citep{Cummings1987, Heber1996a, McDonald1997, McKibben1989} radial and latitudinal gradients of GCR protons, $G_{r}(R)$ and $G_{\theta}(R)$, 
%and Heliums $\left(G_{r,p}(R), G_{\theta,p}(R), G_{r,He}(R), G_{\theta,He}(R)\right)$, 
we have to make the assumption that the variations in time and space can be separated. 
%We compare the intensity $J_{U,s}(R,t,r,\theta)$ of species $s$ measured at Ulysses' position at radial distance $r$ and latitude $\theta$, rigidity $R$ and time $t$ averaged over one solar rotation ($\sim$27 days) with $J_{E,s}(R,t,r_E,\theta_E)$, the corresponding intensity detected at Earth by PAMELA at the same rigidity. 
We compared the intensity $J_{U}(R,t,r,\theta)$ of protons measured at Ulysses' position at radial distance $r$, latitude $\theta$, rigidity $R,$ and time $t$ averaged over one solar rotation ($\sim$27 days) with $J_{E}(R,t,r_E,\theta_E)$, the corresponding proton intensity detected at Earth by PAMELA at the same rigidity (Fig.~\ref{grad_orbit}). 
%
%%%%%%%%%%%%%%%%%%%%%%%%%%%%%%%%%%%%%%%%%%%
%%%%%%%%%%%%%%%%%%%%%%%%%%%%%%%%%%%%%%%%%%%
%
In comparison to the analysis in \citet{DeSimone2011}, the Ulysses/KET data were re-investigated; the response functions were updated,
for instance. Additionally, we used here the improved PAMELA proton data obtained from \citet{Adriani2013} through the ASI Science Data Center (ASDC). 
%, the Helium data from \citet{Nikolay2011}.}}

\begin{figure}
 \centering
% \noindent
 \includegraphics[viewport=54 43 542 383,clip, width=\hsize, angle=0]{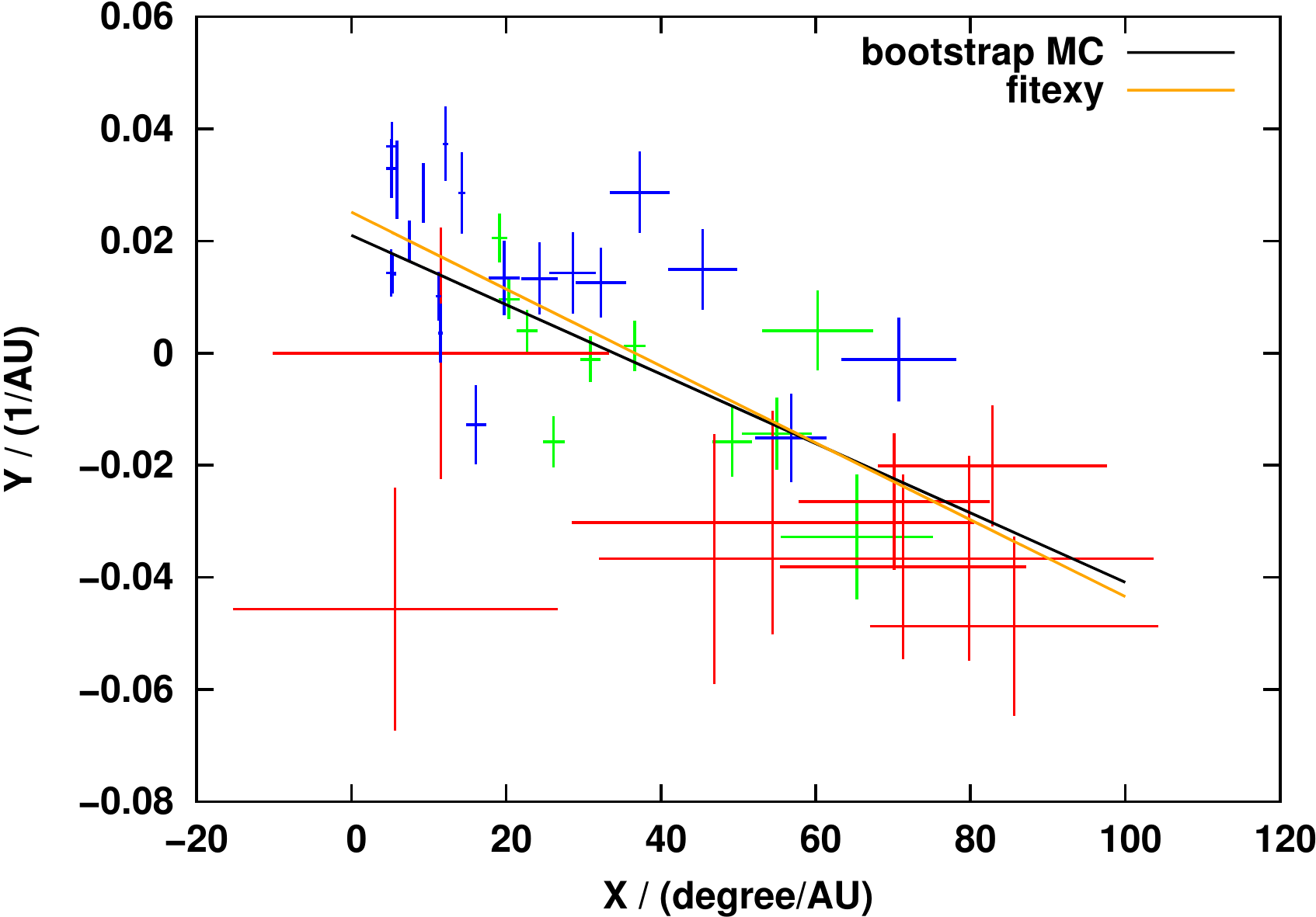}
 \caption{
 $Y$ as a function of $X$ (cf. Eq.~\ref{eq:xy}) for $\sim$1.9\,GV protons.
 %Ulysses/KET P190.3 and corresponding PAMELA channel. 
 See Fig.~\ref{grad_orbit} for color coding. 
 The lines show the results of the two different fit methods, with 
 %a radial and latitudinal gradient of 
 $G_r=(2.14\pm0.21)$\%/AU and $G_\theta=(-0.063\pm0.006)$\%/deg for the bootstrap Monte Carlo approach (black), and $G_r=(2.51\pm0.16)$\%/AU and $G_\theta=(-0.069\pm0.006)$\%/deg for the fit routine using the \texttt{fitexy} algorithm (orange).
 }
 \label{grad_xy}
 \end{figure}

To validate that our measurements from PAMELA and Ulysses/KET are sensitive to the same particle populations, we followed the same approach as described in \citet{DeSimone2011} and investigated the temporal intensity variations of both measurements.
First, we chose two time periods $t_1$ and $t_2$ where Ulysses was at nearly the same radial distance ($r_1 \approx r_2$) and absolute latitude ($|\theta_1| \approx |\theta_2|$). 
%\textcolor{ForestGreen}{These periods in 2006 and 2008 are indicated by shadings in Fig.~\ref{grad_orbit}, where orbit details and selected intensities of PAMELA and Ulysses are plotted over time.}
Then, we built the ratios of intensity spectra measured at $t_1$ and $t_2$ by Ulysses/KET and PAMELA, respectively.
If we assume that the spatial gradients did not change from 2006 to 2009, %over these time periods, 
they cancel out and the ratio measured by Ulysses/KET should be equal to that of PAMELA:
\begin{equation}
%{\frac{J_{U,s}(R, t_1, r_1, \theta_1)}{J_{U,s}(R, t_2, r_2, \theta_2)}} = {\frac{J_{E,s}(R, t_1, r_E, \theta_E)}{J_{E,s}(R, t_2, r_E, \theta_E)}}
{\frac{J_{U}(R, t_1, r_1, \theta_1)}{J_{U}(R, t_2, r_2, \theta_2)}} = {\frac{J_{E}(R, t_1, r_E, \theta_E)}{J_{E}(R, t_2, r_E, \theta_E)}}
\label{eq:ratio_time}
.\end{equation}
This is true for the six proton channels from Ulysses/KET and their corresponding PAMELA channels at comparable rigidities that we used in this study (cf. Table~\ref{table1}).
Each pair of these measurements $J_{U}$ and $J_{E}$ at rigidity $R$
%by Ulysses/KET and PAMELA 
is connected with a function $f(R,\Delta r, \Delta \theta)$ depending on the differences in radial distance and absolute latitude, $\Delta r = r_U - r_E$ and $\Delta\theta = |\theta_U|-|\theta_E|$, respectively:
\begin{equation}
%J_{U,s}(R,t,r,\theta) = J_{E,s}(R,t,r_E,\theta_E) \cdot f(R,\Delta r, \Delta\theta) \label{equation2}
J_{U}(R,t,r,\theta) = J_{E}(R,t,r_E,\theta_E) \cdot f(R,\Delta r, \Delta\theta) \label{equation2}
.\end{equation}
Although some asymmetries have been reported \citep[e.g.][]{Heber1996b,Simpson1996}, we assumed a symmetric distribution of GCRs along the heliographic equator. By separating the variations in radial distance and latitude \citep{Bastian1980, McKibben1979} and approximating them with an exponential function \citep[e.g.][]{Cummings2009}, Eq.~\ref{equation2} can be rewritten as
\begin{equation}
%J_{U,s}(R,t,r,\theta) = J_{E,s}(R,t,r_E,\theta_E) \cdot e^{G_{r,s}\cdot\Delta r} \cdot e^{G_{\theta,s}\cdot\Delta\theta} \label{equation3}
J_{U}(R,t,r,\theta) = J_{E}(R,t,r_E,\theta_E) \cdot e^{G_{r}\cdot\Delta r} \cdot e^{G_{\theta}\cdot\Delta\theta} \label{equation3}
.\end{equation}
To estimate the rigidity-dependent radial and latitudinal gradients $G_{r}(R)$ and $G_{\theta}(R)$, we followed the approach by \citet{Paizis-etal-1995} and further transformed Eq.~\ref{equation3}:
%\begin{equation}
%\ln{\left[\frac{J_U}{J_E}\right]} =  G_r\cdot\Delta r+G_\theta \cdot \Delta\theta\\
%\end{equation}
%\begin{equation}
%\underbrace{\frac{1}{\Delta r}\ln{\left[\frac{J_U}{J_E}\right]}}_{=: Y} =  G_r+G_\theta \cdot \underbrace{\frac{\Delta\theta}{\Delta r}}_{ =: X}\label{eq:xy}\\
%\end{equation}
%\begin{equation}
%Y = G_r+ G_\theta \cdot X
%\end{equation}
%\begin{eqnarray}
%\ln{\left[\frac{J_{U,s}(R)}{J_{E,s}(R)}\right]} &=&  G_{r,s}(R)\cdot\Delta r+G_{\theta,s}(R) \cdot \Delta\theta\\
%\underbrace{\frac{1}{\Delta r}\ln{\left[\frac{J_{U,s}(R)}{J_{E,s}(R)}\right]}}_{=: Y} &=&  G_{r,s}(R)+G_{\theta,s}(R) \cdot \underbrace{\frac{\Delta\theta}{\Delta r}}_{ =: X}\label{eq:xy}\\
%Y_{k}(R) &=& G_{r,s}(R)+ G_{\theta,s}(R) \cdot X\label{eq:xy2}
%\end{eqnarray}
\begin{eqnarray}
\ln{\left[\frac{J_{U}(R)}{J_{E}(R)}\right]} &=&  G_{r}(R)\cdot\Delta r+G_{\theta}(R) \cdot \Delta\theta\\
\underbrace{\frac{1}{\Delta r}\ln{\left[\frac{J_{U}(R)}{J_{E}(R)}\right]}}_{=: Y} &=&  G_{r}(R)+G_{\theta}(R) \cdot \underbrace{\frac{\Delta\theta}{\Delta r}}_{ =: X}\label{eq:xy}\\
Y(R) &=& G_{r}(R)+ G_{\theta}(R) \cdot X\label{eq:xy2}
-\end{eqnarray}
If we assume that the radial and latitudinal gradients 
are constant over the observed time interval and in space,
%are independent of time and space, 
we can calculate their values from the slope and offset by fitting a straight line to the data, as shown for example in Fig.~\ref{grad_xy}, where $Y=\ln{\left[{J_{U}}/{J_{E}}\right]}/{\Delta r}$ is plotted with respect to $X={\Delta\theta}/{\Delta r}$. 
%An example of this procedure is presented in Fig.~\ref{grad_xy} for $\sim$1.9\,GV protons.
% 
%%%%%%%%%%%%%%%%%%%%%%%%%%%%%%%%%%%%%%%%%%%
%%%%%%%%%%%%%%%%%%%%%%%%%%%%%%%%%%%%%%%%%%%

In this figure three different phases of Ulysses' orbit (cf.~Fig.~\ref{grad_orbit}) are indicated by different colors: red shows the fast latitude scan, green and blue indicate the slow ascent and descent over the two solar poles, respectively. The large uncertainties in $X$ and $Y$ for the fast latitude scan data points originate in the wide latitude ranges and the small radial distances, respectively, during this period. 
%[Please add this to figure 2 (Fig.~\ref{grad_orbit}?) showing the rates and orbit.]
We therefore omitted the two data points of Ulysses' closest approach to Earth (the two red data points far left in Fig.~\ref{grad_xy}) in our analysis.
In addition, we estimated the gradients not only for the whole time period, but also separately for the slow ascent (including the first two fast latitude scan data points) and for the slow descent (including the last three fast latitude scan data points) (cf. Table~\ref{table2}).

%\subsection{Fit methods}
%\label{subsection_fits}
Because there are uncertainties in the $X$ and $Y$ dimension, we used two different methods to calculate the fits according to Eq.~\ref{eq:xy2}  including the uncertainties:
 \begin{enumerate}
 \item Fit the data by minimizing the sum of squares using the $\chi^2$ from the \texttt{fitexy} function from Numerical Recipes \citep{Press1996}, which includes $\Delta X$ and $\Delta Y$.
 \item Perform a bootstrap Monte Carlo approach where we take one random value inside of its uncertainties for each data point of Fig.~\ref{grad_xy}. Afterward, a standard minimization was applied to the corresponding ensemble of data points. The whole procedure was carried out 100\,000 times, resulting in mean values over all iterations as the gradients, with the standard deviations as their error. 
 \end{enumerate}
%\textcolor{red}{{Our final results for the radial and latitudinal gradients are then estimated by the mean values of these two methods.}}

\subsection{Normalization}
\label{subsection_normalization}
%
%%%%%%%%%%%%%%%%%%%%%%%%%%%%%%%%%%%%%%%%%%%
\begin{figure}
\centering
%\noindent
\includegraphics[width=\hsize]{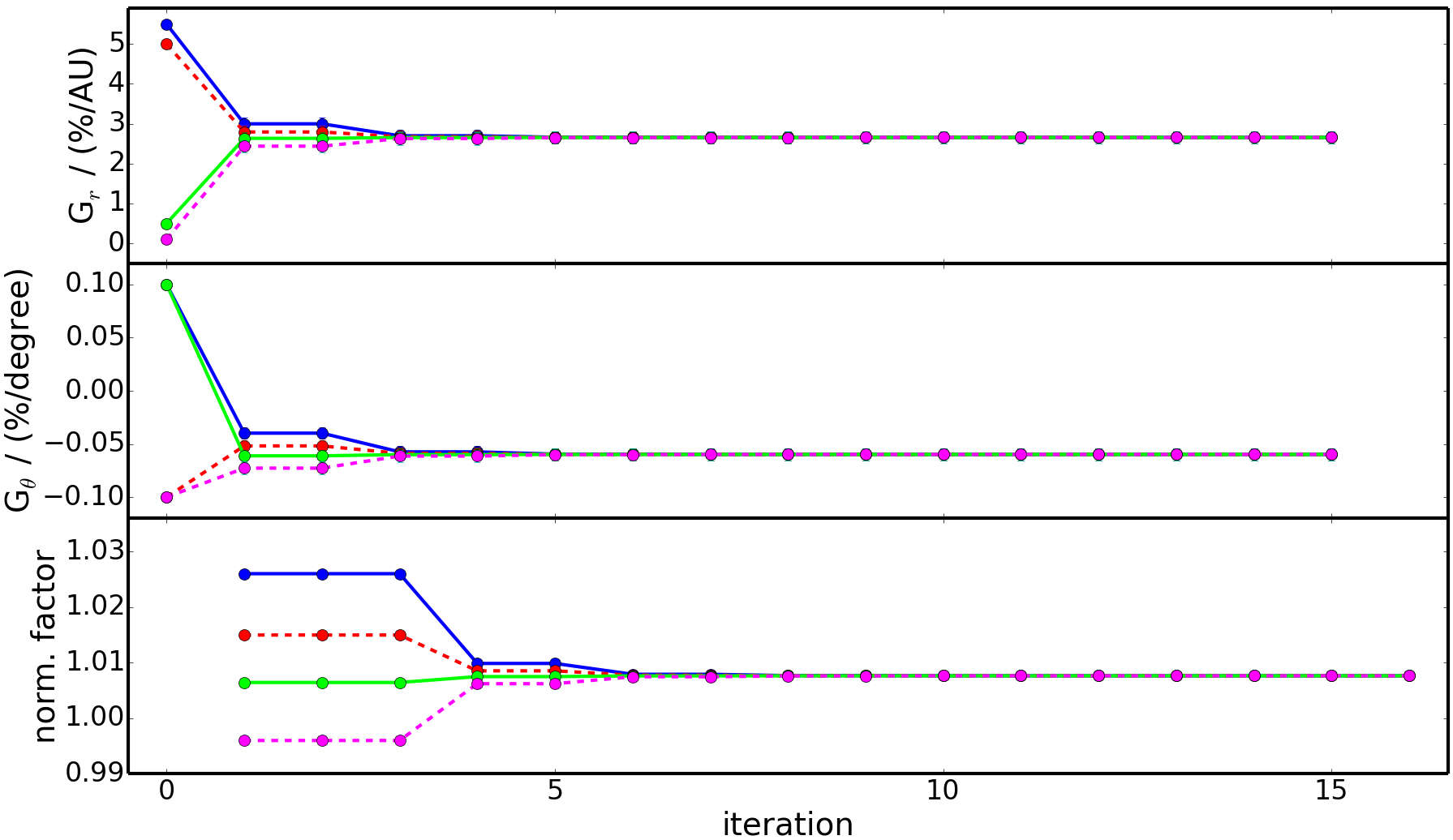}
%\fbox{\includegraphics[height=20pc]{ket_p32_6.png}}
\caption{Iterative normalization of $J_U/J_E$. The top and middle panel show the spatial gradients we used to calculate the normalization factor of $J_U/J_E$ (bottom panel). Shown here are four different iteration runs for $\sim$1.9\,GV protons using the \texttt{fitexy} fit (cf. Sect.~\ref{subsection_normalization} for more details).}
\label{ket_p190_3_norm_iter_comparison}
\end{figure}
%%%%%%%%%%%%%%%%%%%%%%%%%%%%%%%%%%%%%%%%%%%
%
%\textcolor{ForestGreen}{}
Because of non-neglectable uncertainties in the calculation of the absolute intensity values measured by Ulysses/KET, we normalized the measured intensity ratios $J_{U}/J_{E}$ using a 25-day measurement interval in August 2007. During this time, Ulysses was closest to Earth (cf. Fig.~\ref{grad_orbit}), resulting in the smallest gradient effects. Following Eq.~\ref{equation3}, we expect for all rigidities a ratio $J_{U}/J_{E}$ close to 1. The normalized value $\langle{J_{U}(R)}/{J_{E}(R)}\rangle_{N}$ at a given rigidity is determined by the following iterative process:
\begin{eqnarray}
%\Big\langle\frac{J_{U,s}(R)}{J_{E,s}(R)}\Big\rangle_{N}^{k} &=&  \frac{1}{n} \sum_{i=1}^{n=25}  e^{G_{r,s}^{k-1}\cdot\Delta r_{i}} \cdot e^{G_{\theta,s}^{k-1}\cdot\Delta\theta_{i}} \label{norm}
\Big\langle\frac{J_{U}(R)}{J_{E}(R)}\Big\rangle_{N}^{k} &=&  \frac{1}{n} \sum_{i=1}^{n=25}  e^{G_{r}^{k-1}\cdot\Delta r_{i}} \cdot e^{G_{\theta}^{k-1}\cdot\Delta\theta_{i}} \label{norm}
\end{eqnarray}
for iteration step $k$ and the 25 daily values of $\Delta r_{i}$ and $\Delta\theta_{i}$. 
We started with extreme but realistic values for $G_{r}^{k=0}$ and $G_{\theta}^{k=0}$ that covered different scenarios (see Fig.~\ref{ket_p190_3_norm_iter_comparison}). With these starting values for the gradients, we first calculated the intensity normalization $\langle{J_{U}(R)}/{J_{E}(R)}\rangle_{N}^{k=1}$ in August 2007 according to Eq.~\ref{norm}. This normalization factor was then inserted in Eq.~\ref{eq:xy} to adjust the intensity ratio: 
\begin{eqnarray}
\underbrace{\frac{1}{\Delta r}\ln{\left[
\Big\langle\frac{J_{U}(R)}{J_{E}(R)}\Big\rangle_{N}^{k=1} \cdot
\frac{J_{U}(R)}{J_{E}(R)}\right]}}_{=: Y} &=&  G_{r}(R)+G_{\theta}(R) \cdot \underbrace{\frac{\Delta\theta}{\Delta r}}_{ =: X}\label{eq:xy_norm}
.\end{eqnarray}
With Eq.~\ref{eq:xy_norm} %this normed intensity ratio 
we can calculate the spatial gradients $G_{r}(R)$ and $G_{\theta}(R)$.
These were then used in Eq.~\ref{norm} as $G_{r}^{k=1}$ and $G_{\theta}^{k=1}$ for the next iteration $k=2$. This process was repeated until the value for the normalization and thus the spatial gradients converged.
%\textcolor{red}{
%This normalization has been done independently for every channel pair listed in Tab.~\ref{table1}, 
Figure~\ref{ket_p190_3_norm_iter_comparison} exemplarily shows the analysis for $\sim$1.9\,GV protons. For every channel pair, this procedure converges after a few iterations, independently of the starting values for the spatial gradients.

\subsection{Rigidity identification}
\label{subsection_chi}
%
%%%%%%%%%%%%%%%%%%%%%%%%%%%%%%%%%%%%%%%%%%%
\begin{figure}
\centering
%\noindent
\includegraphics[width=\hsize]{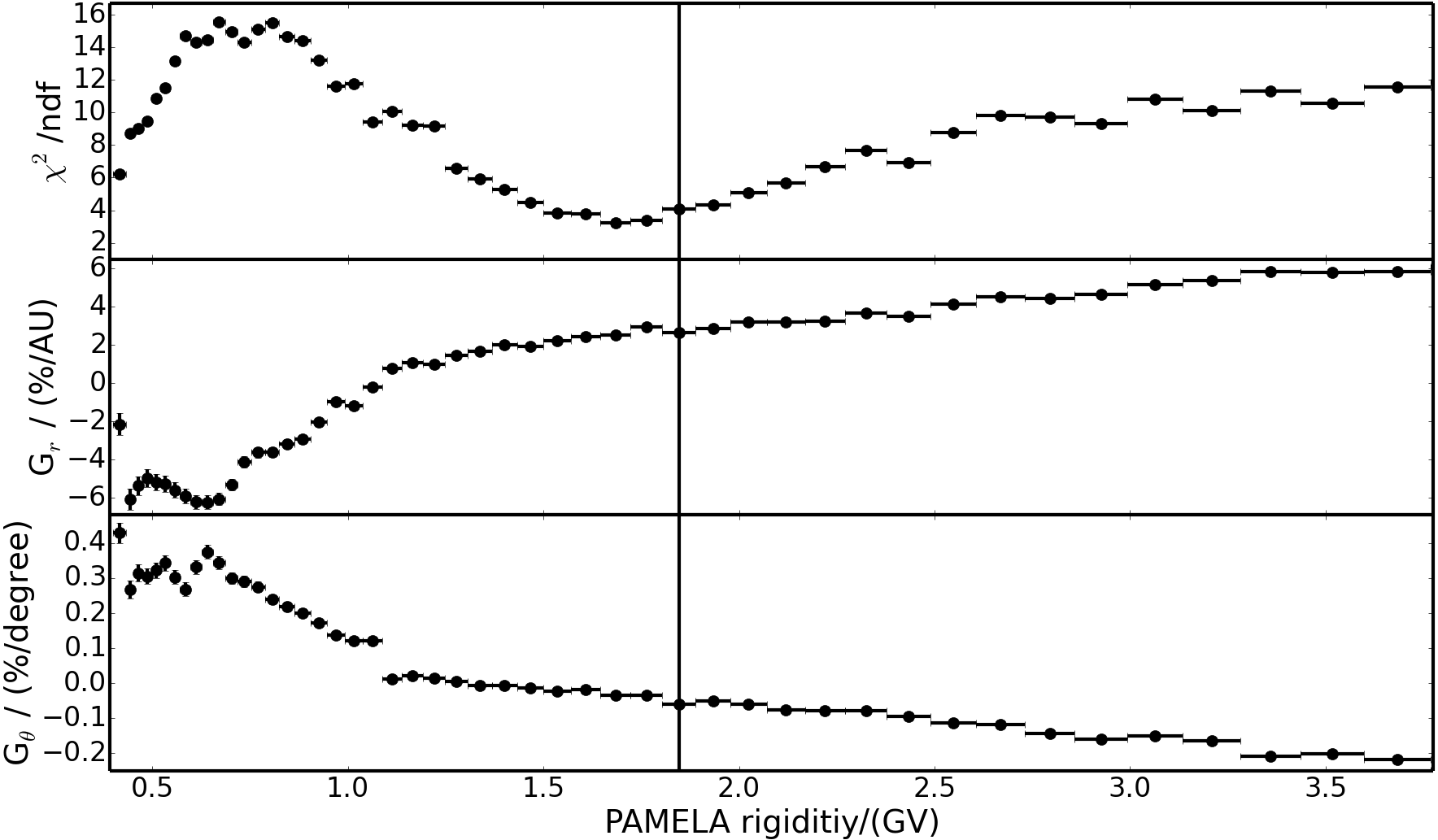}
\caption{
Top: $\chi^2$ of the fit to Eq.~\ref{eq:xy_norm} using the \texttt{fitexy} function %(shown in Fig.~\ref{grad_xy}) 
for protons measured at Ulysses/KET at $\sim$1.9\,GV as a function of PAMELA proton measurements at varying rigidities used in Eq.~\ref{eq:xy_norm}. Middle and bottom: Corresponding radial and latitudinal gradients, respectively. The PAMELA rigidity bin closest to the theoretical mean rigidity of Ulysses/KET is
indicated with the vertical line.%used in our analysis according to Tab.~\ref{table1}.
}
\label{ket_p190_3_chi2}
\end{figure}
%%%%%%%%%%%%%%%%%%%%%%%%%%%%%%%%%%%%%%%%%%%
%
%\textcolor{ForestGreen}{ }
We have already mentioned that the radial and latitudinal gradients are both rigidity dependent. Furthermore, our analysis and especially Eq.~\ref{eq:xy_norm}, from which we obtain the gradients by a fit routine, are only valid if we compare intensities at the same rigidity. Because of this, the quality of the fits is expected to vary if we compare Ulysses/KET measurements at a fixed rigidity with PAMELA measurements at varying rigidities (and vice versa). We used this to check the reliability of our analysis by calculating the different $\chi^2$ of the fits to Eq.~\ref{eq:xy_norm} for PAMELA measurements at varying rigidities. As an example, the results for protons that are measured by Ulysses/KET at $\sim$1.9\,GV are shown in Fig.~\ref{ket_p190_3_chi2} (top panel) together with the resulting radial (middle) and latitudinal (bottom) gradients. 
The PAMELA rigidity bin 
%used in our analysis according to Tab.~\ref{table1}. 
closest to the theoretical mean rigidity of Ulysses/KET is indicated
by the vertical line.
This data point is very close to the absolute $\chi^2$ minimum for the fit, indicating once again that we compare measurements at the same rigidity. In addition, both resulting gradients show only small variations around the rigidity of the $\chi^2$ minimum. The two other proton measurements at rigidities above 1\,GV show similar results. 

However, the three low-rigidity measurements have all the lowest $\chi^2$ at the same comparison rigidity, meaning that they are all sensitive to the same PAMELA rigidity channel. All three have adjacent mean rigidities estimated by Ulysses/KET. If we perform our analysis using the nominal rigidity channels of PAMELA, the corresponding radial and latitudinal gradients show strong variations and are inconsistent in some parts, even within the errors (see Table~\ref{table2}), whereas they agree reasonably
well if we use the very same PAMELA rigidity channel for all three Ulysses/KET channels (see Fig.~\ref{grad_comparison} and Table~\ref{table2}).
This can be mainly attributed to low counting statistics compared to the higher rigidities. We therefore  estimated a mean radial and latitudinal gradient for the three low-rigidity measurements by calculating the arithmetic means of the corresponding gradients that use the same PAMELA channel. %, with standard deviations as errors.

\section{Conclusions}
%
%%%%%%%%%%%%%%%%%%%%%%%%%%%%%%%%%%%%%%%%%%%
 \begin{figure}
\centering
%\noindent
 \includegraphics[viewport=6 99 635 550, clip, width=\hsize, angle=0]{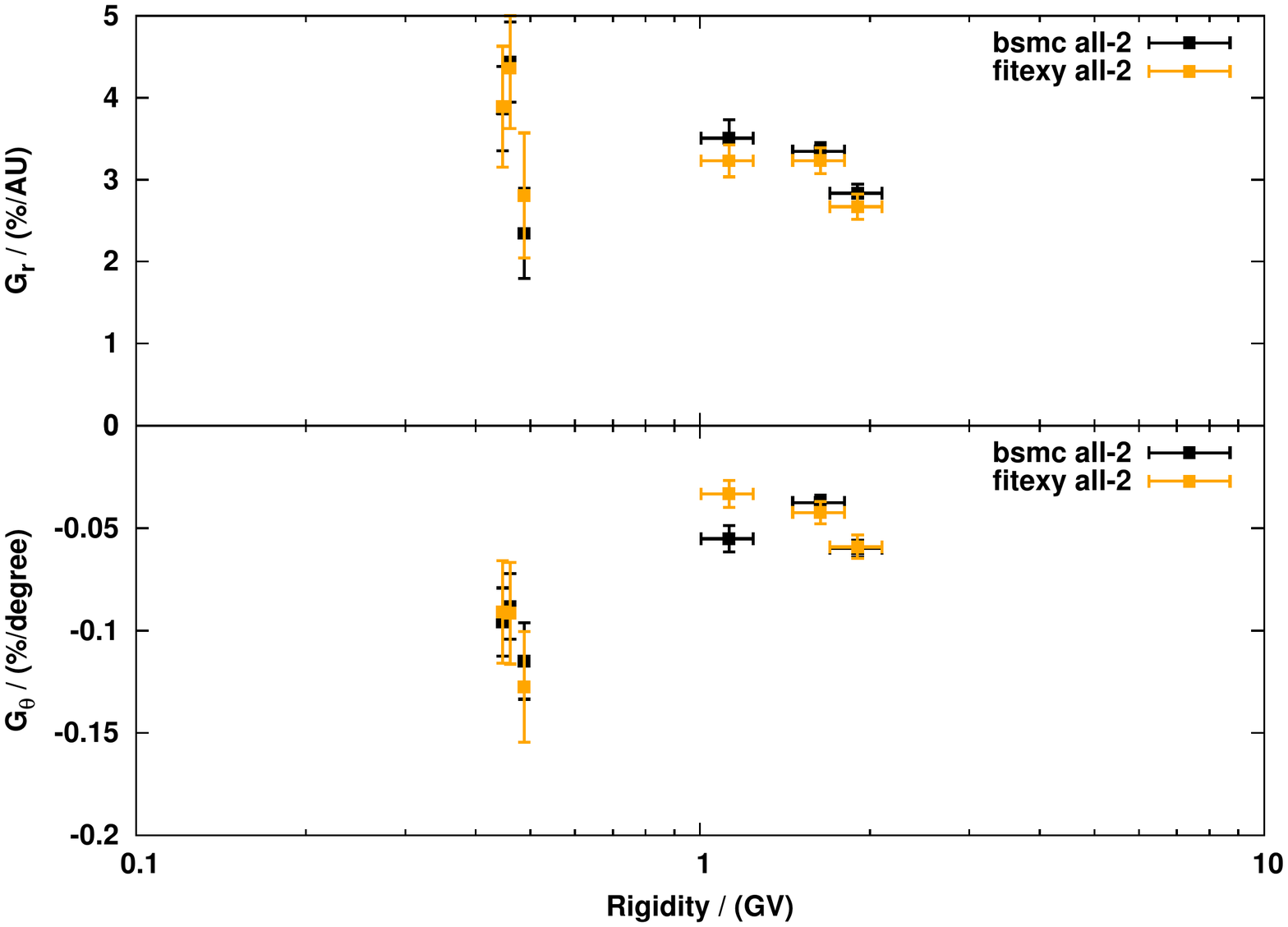}
 \caption{Calculated radial (top) and latitudinal (bottom) gradients for the whole investigation period. Shown are the gradients resulting from the bootstrap Monte Carlo fit approach in black and using the \texttt{fitexy} function in orange. For all three rigidities below 1\,GV the same PAMELA channel is used (see Sect.~\ref{subsection_chi}).}
 \label{grad_comparison}
 \end{figure}
%%%%%%%%%%%%%%%%%%%%%%%%%%%%%%%%%%%%%%%%%%%
%
%
%%%%%%%%%%%%%%%%%%%%%%%%%%%%%%%%%%%%%%%%%%%
\begin{table}
\caption{Proton mean rigidities %$<$$R_U$$>$ and $<$$R_P$$>$ 
as measured by Ulysses/KET and PAMELA that are used to calculate the corresponding gradients. Radial and latitudinal gradients for the whole investigation period, estimated by the arithmetic mean of the corresponding gradients for the two different fit routines (see Table~\ref{table2} for detailed results).}
\label{table1}
\centering
\begin{tabular}{l l l l}
\toprule
%$<$${R_{U}}$$>$/GV & $<$${R_{P}}$$>$/GV & $G_r$ (\%/AU)$^{-1}$ & $G_\theta \cdot ($\%/degree)$^{-1}$ \\
$<$${R_{U}}$$>$/GV & $<$${R_{P}}$$>$/GV & $G_r$\,/\,(\%/AU) & $G_\theta$\,/\,(\%/degree) \\
\midrule
0.46            & 0.46  & 3.6$\, \pm\, $0.7             & -0.10$\, \pm\, $0.03     \\      %arithemtic mean for error (fehlerfortpfl.) - all arith
%0.460  & 0.464 & 3.617$\pm$0.633               & -0.102$\pm$0.022      \\      %arithemtic mean for error (fehlerfortpfl.) - all arith
%0.460  & 0.464 & 3.617$\pm$0.767               & -0.102$\pm$0.015      \\      %arithemtic mean for error (fehlerfortpfl.) ATT: STD bfor
%0.460  & 0.464 & 3.617$\pm$0.069               & -0.102$\pm$0.002      \\      %standard deviation for error
%               &               & 3.548$\pm$0.882               & -0.1$\pm$0.012  \\      %standard deviation for error
%               &               & 3.548$\pm$0.519               & -0.1$\pm$0.018  \\      %arithemtic mean for error (fehlerfortpfl.) 
%               &               & 3.686$\pm$0.651               & -0.103$\pm$0.018        \\      %standard deviation for error
%               &               & 3.686$\pm$0.746               & -0.103$\pm$0.026        \\      %arithemtic mean for error (fehlerfortpfl.) 
%\midrule
1.13            & 1.11  & 3.4$\, \pm\, $0.3             & -0.04$\, \pm\, $0.01     \\      %arithemtic mean for error (fehlerfortpfl.)
%1.125  & 1.113         & 3.369$\pm$0.210               & -0.044$\pm$0.007         \\      %arithemtic mean for error (fehlerfortpfl.)
%1.125  & 1.113         & 3.369$\pm$0.139               & -0.044$\pm$0.011         \\      %standard deviation for error
%               &               & 3.507$\pm$0.224               & -0.055$\pm$0.007        \\
%       &               & 3.23 $\pm$0.196               & -0.033$\pm$0.007         \\
%\midrule
1.63            & 1.61  & 3.3$\, \pm\, $0.2             & -0.04$\, \pm\, $0.01     \\      %arithemtic mean for error (fehlerfortpfl.)
%1.634  & 1.608 & 3.29\ \ $\pm$0.131            & -0.04\ \ $\pm$0.005         \\      %arithemtic mean for error (fehlerfortpfl.)
%1.634  & 1.608 & 3.29\ \ $\pm$0.058            & -0.04\ \ $\pm$0.003         \\      %standard deviation for error 
%               &               & 3.347$\pm$0.104               & -0.037$\pm$0.004        \\ 
%               &               & 3.232$\pm$0.158               & -0.042$\pm$0.006        \\
%\midrule
1.90            & 1.85  & 2.8$\, \pm\, $0.2             & -0.06$\, \pm\, $0.01     \\      %arithemtic mean for error (fehlerfortpfl.)
%1.901  & 1.846 & 2.752$\pm$0.136               & -0.06\ \ $\pm$0.006         \\      %arithemtic mean for error (fehlerfortpfl.)
%1.901  & 1.846 & 2.752$\pm$0.083               & -0.06\ \ $\pm$0.001         \\      %standard deviation for error
%               &               & 2.834$\pm$0.115               & -0.06$\pm$0.005         \\
%               &               & 2.669$\pm$0.157               & -0.059$\pm$0.006        \\
\bottomrule
\end{tabular}
\end{table}
%%%%%%%%%%%%%%%%%%%%%%%%%%%%%%%%%%%%%%%%%%%
%

Our results for the radial and latitudinal gradients for GCR protons during the unusual A$<$0 solar minimum between solar cycle 23 and 24 are summarized in Table~\ref{table1} and Fig.~\ref{grad_comparison} (and in more detail in Table~\ref{table2}). 

The radial gradients vary from $G_r=2.8\pm0.2\%$/AU for 1.9\,GV to  $G_r=3.6\pm0.7\%$/AU for 0.46\,GV, showing values similar to those found in previous studies \citep{Cummings1987,McKibben-etal-1975,Mckibben-1975,Heber1996b}. The values are always positive and show the expected trend of having smaller gradients at higher rigidities. We note that the values agree very well with the one given by \citet{Cummings1987} for the Voyagers.  %In comparison to some previous findings \citep{} we find smaller radial gradients whereas other studies for the same period report similar results \citep{}.

%
%%%%%%%%%%%%%%%%%%%%%%%%%%%%%%%%%%%%%%%%%%%
 \begin{figure}
\centering
%\noindent
 \includegraphics[viewport=54 43 610 379, clip, width=\hsize, angle=0]{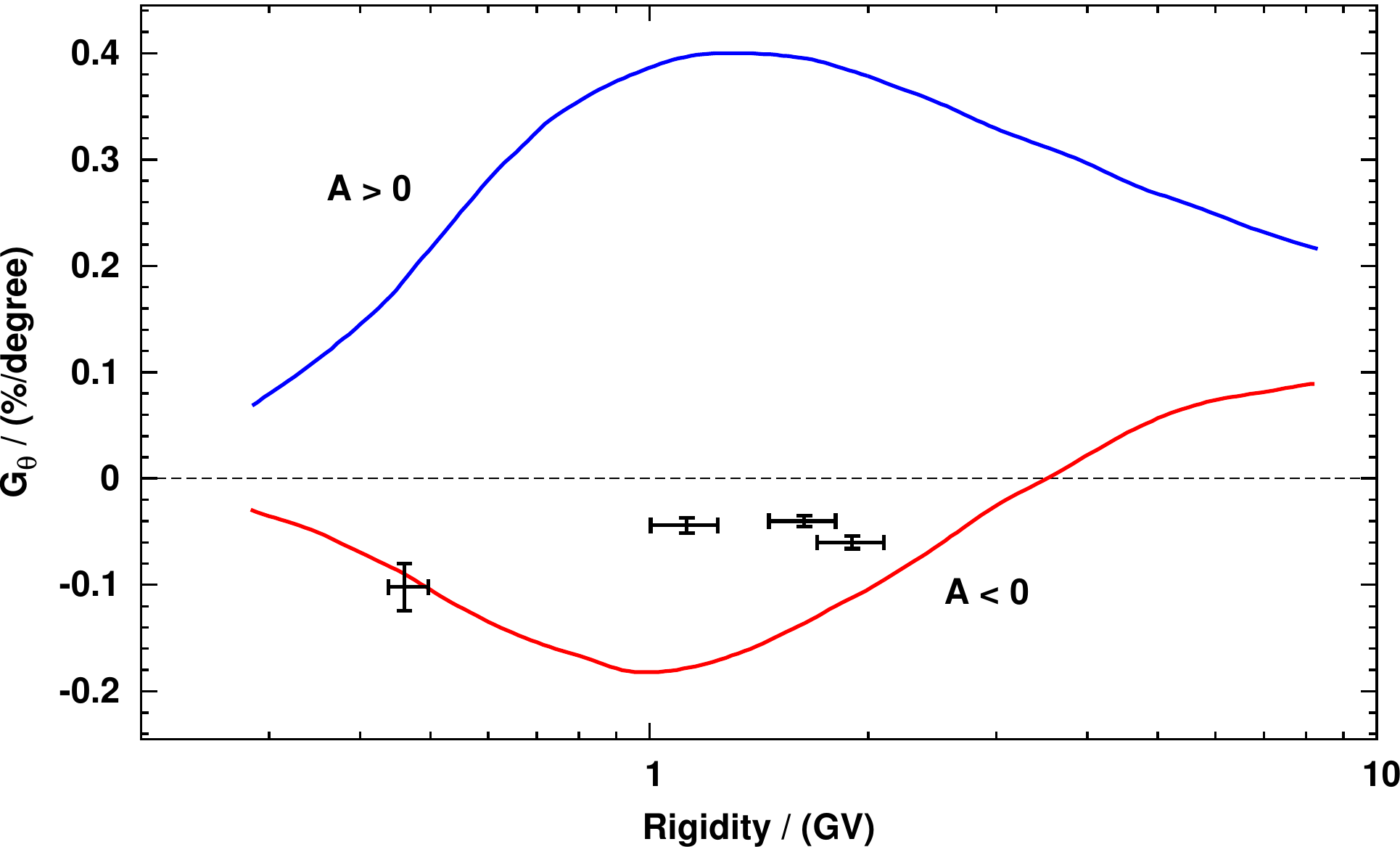}
 \caption{Computed latitudinal gradients for protons during the last A$>$0 solar minimum based on Ulysses/KET measurements (blue line), and a model prediction for the A$<$0 solar minimum investigated here (red line) (figure adapted from \citealt{Potgieter2001}, more details in \citealt{Burger2000}). The mean latitudinal gradients found in this study are plotted in
black.}
 \label{potgieter}
 \end{figure}
%%%%%%%%%%%%%%%%%%%%%%%%%%%%%%%%%%%%%%%%%%%
%
In agreement with the prediction of calculations solving the Parker transport equation (PTE) for an A$<$0-solar magnetic epoch, the measured latitudinal gradients are negative \citep[see, e.g.,][and references therein]{Potgieter2001,Potgieter-etal-2014}, with values from $G_\theta=-0.04\pm0.01\%$/degree for 1.13\,GV and 1.63\,GV to $G_\theta=-0.1\pm0.03\%$/degree for 0.46\,GV.
In contrast to the radial gradients, the latitudinal gradient for protons is much smaller than the values reported by \citet{Cummings1987} for the last A$<$0-solar magnetic epoch. 
The absolute magnitude, however, is smaller for protons above 1\,GV and significantly larger for protons below 0.5\,GV if compared to Ulysses/KET measurements during the declining phase of solar cycle 23 \citep[Fig.~7 in][]{Heber1996a}. 
Updated data sets and minor changes in the analysis procedures allowed us to obtain slightly larger latitudinal gradients than \citet{DeSimone2011}, who reported a gradient $G_{\theta}=-0.024\pm0.005\%$/deg for $\sim$1.7\,GV protons. 

\citet{Potgieter2001} performed calculations solving the PTE by adapting the transport parameters in a way that the rigidity dependence of the maximum latitudinal gradient measured by Ulysses/KET during the declining phase of solar cycle 23 \citep[Fig.~7 in][]{Heber1996a} is well reproduced. 
We note that this measured maximum latitudinal gradient shows the same rigidity dependency as the mean latitudinal gradient with only a constant offset.
The results of these calculations are summarized by the blue curve in Fig.~\ref{potgieter}. Assuming that Ulysses were to perform its third orbit under similar conditions as the first orbit but with opposite solar magnetic polarity, the authors found a rigidity dependence of the latitudinal gradient as given by the red curve in Fig.~\ref{potgieter}. 
It is important to note that in agreement with our measurements, the predictions give lower values for the A$<$0-solar magnetic epoch and a maximum of the gradient towards lower rigidities. In contrast to the prediction, this maximum either does not exist or a maximum is present at even lower rigidities. Because the gradients at rigidities above 1\,GV do not resemble the predicted rigidity dependence, it seems to us unlikely that the gradient may become positive at even higher values. 

Since the very local interstellar spectrum has been determined by recent Voyager, PAMELA, and AMS observations \citep{Potgieter-etal-2014}, the results reported here are crucial for evaluating the transport parameters in the heliosphere. 

\begin{acknowledgements}
The Ulysses/KET project is supported under Grant 50 OC 1302 by the German Bundesministerium f\"ur Wirtschaft through the Deutsches Zentrum f\"ur Luft- und Raumfahrt (DLR). 
%The PAMELA mission is sponsored by the Italian National Institute of Nuclear Physics (INFN), the Italian Space Agency (ASI), the Russian Space Agency (Roskosmos), the Russian Academy of Science, the Deutsches Zentrum f\"ur Luft- und Raumfahrt (DLR), the Swedish National Space Board (SNSB) and the Swedish Research Council (VR). 
%\textcolor{red}{The German-Italian collaboration has been supported by the Deutsche Forschungsgemeinschaft under grant HE3279/11-1.}
PAMELA proton data obtained via the web site http://tools.asdc.asi.it/cosmicRays.jsp from the Italian Space Agency (ASI) Science Data Center.
Ulysses and PAMELA orbit data obtained via the web site http://omniweb.gsfc.nasa.gov/coho/helios/heli.html from NASA's Space Physics Data Facility (SPDF). 
We acknowledge the NMDB database (www.nmdb.eu), founded under the European Union's FP7 program (contract no. 213007) for providing the Kiel neutron monitor data.
Sunspot number data used in this study was obtained via the web site http://sidc.be/sunspot-data/ courtesy of the SIDC-team, World Data Center for the Sunspot Index, Royal Observatory of Belgium.
Wilcox Solar Observatory data used in this study was obtained via the web site http://wso.stanford.edu courtesy of J.T. Hoeksema. 
\end{acknowledgements}

% WARNING
%-------------------------------------------------------------------
% Please note that we have included the references to the file aa.dem in
% order to compile it, but we ask you to:
%
% - use BibTeX with the regular commands:
%   \bibliographystyle{aa} % style aa.bst
%   \bibliography{Yourfile} % your references Yourfile.bib
%
% - join the .bib files when you upload your source files
%-------------------------------------------------------------------

\bibliographystyle{aa} % style aa.bst
\bibliography{references} % your references.bib

%\begin{thebibliography}{}
%\end{thebibliography}

\begin{appendix}
\section{Detailed gradient results}
%
%%%%%%%%%%%%%%%%%%%%%%%%%%%%%%%%%%%%%%%%%%%
\begin{table*}%[!t]%[ht]
%\small
%\footnotesize
%\begin{center}
%\tabcolsep=0.1cm
%\hspace{-2cm}
%\makebox[\textwidth]{
\caption{Proton mean rigidities $<$$R_U$$>$ and $<$$R_P$$>$ (both in GV) as measured by Ulysses/KET and PAMELA, respectively, that are used to calculate the corresponding gradients. Radial and latitudinal gradients (in \%/AU and \%/degree, respectively), and quality of fit for different selection criteria (cf. Fig.~\ref{grad_orbit}) and fit methods. Indicated by (1) are the values obtained by the bootstrap Monte Carlo approach, while (2) marks the fit routine using the \texttt{fitexy} algorithm. 
Note that each gradient in the last row is calculated by the arithmetic mean of the corresponding gradients of rows 2, 3, and 4
%, with the standard deviation as error 
(see Sect.~\ref{subsection_chi} for details). 
%See Tab.~\ref{table1} for corresponding rigidities.
}
\label{table2}
\centering      
{
\begin{tabular}{@{}l l r l l l l l r@{}}%{|c||c|c|c|c|c|c|c|c|}
\toprule 
% & \multicolumn{2}{c}{all data (-2)} & \multicolumn{2}{c}{slow ascent (green+2)} & \multicolumn{2}{c}{slow descent (blue+3)}\\
${<}$${\textit{R}}_{{\textit{U}}}$${>}$ & ${<}$${\textit{R}}_{{\textit{P}}}$${>}$ & & \multicolumn{2}{c}{all data} & \multicolumn{2}{c}{slow ascent (green)} & \multicolumn{2}{c}{slow descent (blue)}\\
 \cmidrule(lr){4-5} \cmidrule(lr){6-7} \cmidrule(lr){8-9}
 &  &  & \multicolumn{1}{c}{Fit (1)} & \multicolumn{1}{c}{Fit (2)} & \multicolumn{1}{c}{Fit (1)} & \multicolumn{1}{c}{Fit (2)} & \multicolumn{1}{c}{Fit (1)} & \multicolumn{1}{c}{Fit (2)}\\ 
\midrule
{0.45} & \multicolumn{1}{l}{{0.44}}\\
 &  &  $G_r=$ & 3.8$\pm$0.7  & 3.3$\pm$0.9 & 5.4$\pm$1.5 & 3.5$\pm$2.2 & 3.5$\pm$0.7 & \multicolumn{1}{l}{3.2$\pm$0.9}\\
% \multirow{2}{*}{$G_\theta$} & -0.170 & -0.138 & -0.169 & -0.128 & -0.166 & \multicolumn{1}{l}{-0.139}\\
% & \multicolumn{1}{r}{$\pm$0.02} & \multicolumn{1}{r}{$\pm$0.029} & \multicolumn{1}{r}{$\pm$0.04}& \multicolumn{1}{r}{$\pm$0.063} & \multicolumn{1}{r}{$\pm$0.024} & $\pm$0.037\\         
 &  & $G_\theta=$ & -0.17$\pm$0.02 & -0.14$\pm$0.03 & -0.17$\pm$0.04 & -0.13$\pm$0.07 & -0.17$\pm$0.03 & \multicolumn{1}{l}{-0.14$\pm$0.04}\\ 
 &  &  $\chi^2$/ndf = & 0.8 & 0.7 & 0.5 & 0.4 & 1. & \multicolumn{1}{l}{1.}\\ 
\midrule
{0.45} & \multicolumn{1}{l}{{0.46}}\\
 &  & $G_r=$ & 3.9$\pm$0.6  & 3.9$\pm$0.8  & 4.2$\pm$1.2  & 3.6$\pm$1.9  & 3.8$\pm$0.6 & \multicolumn{1}{l}{3.9$\pm$0.8}\\
 &  & $G_\theta=$ & -0.01$\pm$0.02 & -0.09$\pm$0.03 & -0.09$\pm$0.04 & -0.08$\pm$0.06 & -0.1$\pm$0.03 & \multicolumn{1}{l}{-0.09$\pm$0.04}\\
 &  & $\chi^2$/ndf = & 0.6 & 0.6 & 0.3 & 0.2 & 0.9 & \multicolumn{1}{l}{0.9}\\ 
\midrule
{0.46} & \multicolumn{1}{l}{{0.46}}\\
 &  &  $G_r=$ & 4.4$\pm$0.5 & 4.4$\pm$0.8 & 4.4$\pm$1.2 & 3.8$\pm$2.  & 4.3$\pm$0.6 & \multicolumn{1}{l}{4.5$\pm$0.8}\\
% \multirow{2}{*}{$G_\theta$} & -0.088 & -0.091 & -0.092 & -0.08 & -0.095 & \multicolumn{1}{l}{-0.093}\\
% & \multicolumn{1}{r}{$\pm$0.016} & \multicolumn{1}{r}{$\pm$0.025} & \multicolumn{1}{r}{$\pm$0.033}& \multicolumn{1}{r}{$\pm$0.054} & \multicolumn{1}{r}{$\pm$0.02} & $\pm$0.032\\  
 &  & $G_\theta=$ & -0.09$\pm$0.02 & -0.09$\pm$0.03 & -0.09$\pm$0.04 & -0.08$\pm$0.06 & -0.1$\pm$0.02 & \multicolumn{1}{l}{-0.09$\pm$0.04}\\
 &  & $\chi^2$/ndf = & 0.8 & 0.8 & 0.3 & 0.3 & 1.2 & \multicolumn{1}{l}{1.2}\\ 
%\midrule
%{0.460} & \multicolumn{1}{l}{{0.464}}\\
% &  $G_r$ & 4.434$\pm$0.49 & 4.360$\pm$0.735 & 4.441$\pm$1.181 & 3.823$\pm$1.91 & 4.350$\pm$0.511 & \multicolumn{1}{l}{4.451$\pm$0.8}\\
% & $G_\theta$ & -0.088$\pm$0.016 & -0.092$\pm$0.025 & -0.092$\pm$0.033 & -0.08$\pm$0.054 & -0.095$\pm$0.02 & \multicolumn{1}{l}{-0.093$\pm$0.032}\\
% & \quad\quad $\chi^2$/ndf & 0.78 & 0.77 & 0.24 & 0.23 & 1.15 & \multicolumn{1}{l}{1.15}\\ 
%
\midrule
{0.49} & \multicolumn{1}{l}{{0.46}}\\
 &  &  $G_r=$ & 2.3$\pm$0.6  & 2.8$\pm$0.8  & 2.5$\pm$1.3  & 2.8$\pm$2.1  & 2.4$\pm$0.6  & \multicolumn{1}{l}{2.8$\pm$0.9}\\
 &  & $G_\theta=$ & -0.12$\pm$0.02 & -0.13$\pm$0.03 & -0.13$\pm$0.04 & -0.14$\pm$0.06 & -0.13$\pm$0.03 & \multicolumn{1}{l}{-0.13$\pm$0.04}\\
 &  &  $\chi^2$/ndf = & 0.8 & 0.8 & 0.3 & 0.3 & 1.2 & \multicolumn{1}{l}{1.1}\\ 
\midrule
{0.49} & \multicolumn{1}{l}{{0.49}}\\
 &  &  $G_r=$ & 1.6$\pm$0.6 & 2.7$\pm$0.8 & 1.8$\pm$1.3 & 2.9$\pm$2. & 1.6$\pm$0.6  & \multicolumn{1}{l}{2.7$\pm$0.9}\\
% \multirow{2}{*}{$G_\theta$} & -0.093 & -0.124 & -0.113 & -0.137 & -0.103 & \multicolumn{1}{l}{-0.121}\\
% & \multicolumn{1}{r}{$\pm$0.018} & \multicolumn{1}{r}{$\pm$0.027} & \multicolumn{1}{r}{$\pm$0.036}& \multicolumn{1}{r}{$\pm$0.059} & \multicolumn{1}{r}{$\pm$0.021} & $\pm$0.034\\ 
 &  & $G_\theta=$ & -0.09$\pm$0.02 & -0.12$\pm$0.03 & -0.11$\pm$0.04 & -0.14$\pm$0.06 & -0.10$\pm$0.03 & \multicolumn{1}{l}{-0.12$\pm$0.04}\\
 &  &  $\chi^2$/ndf = & 1.1 & 1. & 0.6 & 0.6 & 1.5 & \multicolumn{1}{l}{1.4}\\ 
\midrule
{1.13} & \multicolumn{1}{l}{{1.11}}\\
 &   & $G_r=$ & 3.5$\pm$0.3 & 3.2$\pm$0.2  & 4.9$\pm$0.4        & 5.3$\pm$0.6 & 3.2$\pm$0.2 & \multicolumn{1}{l}{2.8$\pm$0.3}\\
% \multirow{2}{*}{$G_\theta$} & -0.055 & -0.033 & -0.037 & -0.051 & -0.061 & \multicolumn{1}{l}{-0.045}\\
% & \multicolumn{1}{r}{$\pm$0.007} & \multicolumn{1}{r}{$\pm$0.007} & \multicolumn{1}{r}{$\pm$0.01}& \multicolumn{1}{r}{$\pm$0.017} & \multicolumn{1}{r}{$\pm$0.006} & $\pm$0.009\\ 
%
 & &  $G_\theta=$ & -0.06$\pm$0.01 & -0.03$\pm$0.01 & -0.04$\pm$0.01 & -0.05$\pm$0.02 & -0.06$\pm$0.01 & \multicolumn{1}{l}{-0.05$\pm$0.01}\\
 &  &  $\chi^2$/ndf = & 5.3 & 5. & 1.8 & 1.8 & 5. & \multicolumn{1}{l}{4.8}\\ 
\midrule
{1.63} & \multicolumn{1}{l}{{1.61}}\\
 &  &  $G_r=$ & 3.3$\pm$0.2 & 3.2$\pm$0.2 & 2.2$\pm$0.3 & 2.2$\pm$0.4 & 3.6$\pm$0.2   & \multicolumn{1}{l}{3.4$\pm$0.2}\\ 
% \multirow{2}{*}{$G_\theta$} & -0.037 & -0.042 & -0.028 & -0.031 & -0.039 & \multicolumn{1}{l}{-0.034}\\
% & \multicolumn{1}{r}{$\pm$0.004} & \multicolumn{1}{r}{$\pm$0.006} & \multicolumn{1}{r}{$\pm$0.007}& \multicolumn{1}{r}{$\pm$0.011} & \multicolumn{1}{r}{$\pm$0.005} & $\pm$0.007\\ 
%
 & &  $G_\theta=$ & -0.04$\pm$0.01 & -0.04$\pm$0.01 & -0.03$\pm$0.01 & -0.03$\pm$0.02 & -0.04$\pm$0.01 & \multicolumn{1}{l}{-0.03$\pm$0.01}\\
 & &  $\chi^2$/ndf = & 5.1 & 4.9 & 3.6 & 3.6 & 5.7 & \multicolumn{1}{l}{5.7}\\ 
\midrule
{1.90} & \multicolumn{1}{l}{{1.85}}\\
 &   & $G_r=$ & 2.8$\pm$0.2 & 2.7$\pm$0.2 & 2.1$\pm$0.3 & 2.3$\pm$0.4 & 3.1$\pm$0.2  & \multicolumn{1}{l}{2.7$\pm$0.2}\\
% \multirow{2}{*}{$G_\theta$} & -0.06 & -0.059 & -0.059 & -0.064 & -0.053 & \multicolumn{1}{l}{-0.041}\\ 
% & \multicolumn{1}{r}{$\pm$0.005} & \multicolumn{1}{r}{$\pm$0.006} & \multicolumn{1}{r}{$\pm$0.008}& \multicolumn{1}{r}{$\pm$0.012} & \multicolumn{1}{r}{$\pm$0.005} & $\pm$0.007\\ 
%
 &  & $G_\theta=$ & -0.06$\pm$0.01 & -0.06$\pm$0.01 & -0.06$\pm$0.01 & -0.06$\pm$0.02 & -0.05$\pm$0.01 & \multicolumn{1}{l}{-0.04$\pm$0.01}\\
 &  &  $\chi^2$/ndf = & 4.5 & 4.4 & 3.7 & 3.7 & 4.5 & \multicolumn{1}{l}{4.2}\\ 
\midrule
%{0.460}$^\textbf{\textdagger}$ & \multicolumn{1}{l}{{0.464}}\\
% &  & $G_r=$ & 3.548$\pm$0.882 & 3.686$\pm$0.651 & 3.696$\pm$0.864 & 3.423$\pm$0.428  & 3.491$\pm$0.836 & \multicolumn{1}{l}{3.709$\pm$0.696}\\
% &  & $G_\theta=$ & -0.1$\pm$0.012 & -0.103$\pm$0.018 & -0.105$\pm$0.019 & -0.101$\pm$0.027 & -0.108$\pm$0.015 & \multicolumn{1}{l}{-0.103$\pm$0.016}\\                                                  
%%\quad\quad $\chi^2$/ndf & 0.78 & 0.77 & 0.24 & 0.23 & 1.15 & \multicolumn{1}{l}{1.15}\\ 
%%
%\midrule
{0.46}$^\textbf{\textdagger}$ & \multicolumn{1}{l}{{0.46}}\\
 &  & $G_r=$ & 3.5$\pm$0.6      & 3.7$\pm$0.8 & 3.7$\pm$1.3 & 3.4$\pm$2.  & 3.5$\pm$0.6 & \multicolumn{1}{l}{3.7$\pm$0.9}\\
 &  & $G_\theta=$ & -0.1$\pm$0.02 & -0.10$\pm$0.03 & -0.11$\pm$0.04 & -0.10$\pm$0.06 & -0.11$\pm$0.03 & \multicolumn{1}{l}{-0.10$\pm$0.04}\\
\bottomrule
\addlinespace[.5em]
\multicolumn{9}{l}{{\textdagger} mean radial and latitudinal gradients for the three low-rigidity measurements (rows 2, 3, and 4)}
\end{tabular}
}
%\end{center}
\end{table*}
%%%%%%%%%%%%%%%%%%%%%%%%%%%%%%%%%%%%%%%%%%%
%
\end{appendix}

\end{document}